\documentclass[sigconf,final]{acmart}

\usepackage{booktabs}
\usepackage{amsmath}
\usepackage{algorithm}
\usepackage{algorithmic}
\usepackage{graphicx}
\usepackage{subcaption}
\usepackage{hyperref}

\setcopyright{cc}
\setcctype{by}
\copyrightyear{2025}
\acmYear{2025}
\acmConference[ICAIF '25]{6th ACM International Conference on AI in Finance}{November 15--18, 2025}{Singapore, Singapore}
\acmBooktitle{6th ACM International Conference on AI in Finance (ICAIF '25), November 15--18, 2025, Singapore, Singapore}
\acmDOI{10.1145/3768292.3770426}
\acmISBN{979-8-4007-2220-2/2025/11}

\begin{document}

\title{Learning to Manage Investment Portfolios beyond Simple Utility Functions}

\author{Maarten P. Scholl}
\email{maarten.scholl@st-hughs.ox.ac.uk}
\authornote{Corresponding Author}
\affiliation{%
  \institution{University of Oxford} % Department of Computer Science, Oxford Martin School
  \city{Oxford}
  \country{United Kingdom}
}
\author{Mahmoud Mahfouz}
\email{mahmoud.mahfouz@jpmorgan.com} 
\affiliation{%
  \institution{J.P. Morgan Chase \& Co}
  \city{London}
  \country{United Kingdom}
}
\author{Anisoara Calinescu}
\email{anisoara.calinescu@cs.ox.ac.uk}
\affiliation{%
  \institution{University of Oxford}
  \city{Oxford}
  \country{United Kingdom}
}
\author{J. Doyne Farmer}
\email{doyne.farmer@inet.ox.ac.uk}
\affiliation{%
  \institution{University of Oxford}
  \city{Oxford}
  \country{United Kingdom}
}

\begin{abstract}
    While investment funds publicly disclose their objectives in broad terms, their managers optimize for complex combinations of competing goals that go beyond simple risk-return trade-offs. Traditional approaches attempt to model this through multi-objective utility functions, but face fundamental challenges in specification and parameterization. We propose a generative framework that learns latent representations of fund manager strategies without requiring explicit utility specification.

    Our approach directly models the conditional probability of a fund's portfolio weights, given stock characteristics, historical returns, previous weights, and a latent variable representing the fund's strategy. Unlike methods based on reinforcement learning or imitation learning, which require specified rewards or labeled expert objectives, our GAN-based architecture learns directly from the joint distribution of observed holdings and market data.
    
    We validate our framework on a dataset of 1436 U.S. equity mutual funds. The learned representations successfully capture known investment styles, such as "growth" and "value," while also revealing implicit manager objectives. For instance, we find that while many funds exhibit characteristics of Markowitz-like optimization, they do so with heterogeneous realizations for turnover, concentration, and latent factors. %  that are difficult to specify explicitly 
    To analyze and interpret the end-to-end model, we develop a series of tests that explain the model, and we show that the benchmark's expert labeling are contained in our model's encoding in a linear interpretable way.
    Our framework provides a data-driven approach for characterizing investment strategies for applications in market simulation, strategy attribution, and regulatory oversight.
\end{abstract}

\begin{CCSXML}
    <ccs2012>
       <concept>
           <concept_id>10010147.10010257.10010293.10010294</concept_id>
           <concept_desc>Computing methodologies~Neural networks</concept_desc>
           <concept_significance>500</concept_significance>
           </concept>
       <concept>
           <concept_id>10003752.10003790.10003795</concept_id>
           <concept_desc>Applied computing~Economics</concept_desc>
           <concept_significance>300</concept_significance>
           </concept>
     </ccs2012>
\end{CCSXML}

\ccsdesc[500]{Computing methodologies~Neural networks}
\ccsdesc[300]{Applied computing~Economics}

\keywords{generative adversarial networks, portfolio management, investment strategies, imitation learning, agent-based modeling}

\maketitle

%%%%%%%%%%%%%%%%%%%%%%%%%%%%%%%%%%%%%%%%%%%%%%%%%%%%%%%%%%%%%%%%%%%%%%%%%%%%%%%%%%%%%%%%%%%%%%%%%%%%
% 1 INTRODUCTION

\section{Introduction}

Modern Portfolio Theory assumes fund managers maximize a simple utility function that balances risk and return \cite{Markowitz1952}. 
In practice, managers balance multiple competing objectives: tracking error limits, turnover costs, liquidity requirements, regulatory mandates, and behavioral biases \cite{Sharpe1966MutualPerformance}. While these could in theory be combined into complex utility functions, this approach faces the fundamental challenge of specifying unknown objective weights that vary across managers, time, and economic regimes.
Generative modeling offers new ways to understand complex behaviors without specifying explicit objectives. Generative Adversarial Networks (GANs) are frequently used to generate synthetic financial time series \cite{Cont2022TailGAN}, but no one has applied them \emph{to learn investment strategies and generate realistic populations of investors}. This gap matters for agent-based market simulations, which need realistic models of diverse market participants \cite{farmer2009economy}.

We present a generative adversarial framework that sidesteps utility specification problems by learning fund manager strategies directly from portfolio holdings data. We model a manager's strategy as a conditional probability distribution over portfolio weights, without specifying a utility function. Our goal is not theoretical optimality of the portfolio, but capturing real-world manager behavior with all its complexity and imperfections.

Our method enables three applications. First, strategy discovery: the model learns representations that capture known style factors like "value" and "growth" alongside subtle, implicit objectives. Second, behavioral cloning: the model generates realistic portfolio allocations for any market state. These synthetic portfolios enable stress testing and counterfactual analysis. Third, agent-based modeling: our framework creates diverse, realistic agents for market simulations.

% We formulate strategy learning as conditional generation of portfolio allocation: $p\left(\mathbf{w}_t \mid \mathbf{X}, \mathbf{r}, \boldsymbol{\phi}, \mathbf{w}_{t{-}1}\right)$. We design a generative architecture that combines financial factor models with adversarial training. We comprehensively evaluate our framework on 1436 U.S. mutual fund portfolios, showing superior performance over baselines. Finally, we provide evidence that learned representations capture known financial patterns while revealing undisclosed manager objectives, such as turnover rate constraints and mean-variance optimization.
We make four contributions. 

% \begin{enumerate}
(1) Formulating strategy learning as conditional generation of portfolio allocation:
      \( p\left(\mathbf{w}_t \mid \mathbf{X}, \mathbf{r}, \boldsymbol{\phi}, \mathbf{w}_{t-1}\right) \).
      
(2) Designing a generative architecture that integrates financial factor models with adversarial training.

(3) Evaluating the framework comprehensively on 1,436 U.S. mutual fund portfolios, demonstrating superior performance over baselines.

(4) Providing evidence that learned representations capture known financial patterns and reveal undisclosed manager objectives.
% \end{enumerate}

%%%%%%%%%%%%%%%%%%%%%%%%%%%%%%%%%%%%%%%%%%%%%%%%%%%%%%%%%%%%%%%%%%%%%%%%%%%%%%%%%%%%%%%%%%%%%%%%%%%%
% 2 RELATED WORK
\section{Related Work}

Our research builds on four areas: GANs in finance, strategy classification, imitation learning, and agent-based modeling.

\subsection{GANs in Finance}

Generative Adversarial Networks have gained traction in finance, primarily for synthetic data generation. QuantGAN uses temporal convolutional networks to generate financial price series \cite{Wiese2019QuantSeries}. Conditional tabular GANs have been adapted for financial tasks \cite{Xu2019ModelingGAN}, including portfolio optimization for higher Sharpe ratios \cite{Ramirez2023AAllocation}. 
Recent work has further explored these topics for stylized facts in equities markets \cite{Kwon2024CanSeries} and factor models \citep{Gopal2024NeuralFactors}. Unlike these approaches that focus on either data generation or optimization, our work uses GANs to 
learn and represent underlying investment strategies from portfolio holdings.

\subsection{Strategy Classification}

The task of identifying a fund's investment strategy has traditionally been approached from two main angles. The first, known as returns-based style analysis (RBSA), uses a fund's historical returns to regress them against the returns of various market indices or factors, thereby inferring the fund's style \cite{Sharpe1992AssetAllocation,Hasanhodzic2007CanCase}. The second, holdings-based analysis, examines the characteristics of the securities within a fund's portfolio at a given point in time to classify its strategy \cite{Brown2009StayingFunds}. Recently, machine learning techniques have improved upon these methods. For instance, researchers have applied neural networks to fund characteristics to achieve higher classification accuracy \cite{Kaniel2023Machine-learningManagers}, while others have used machine learning to explore the nonlinear relationships between a fund's holdings and its style \cite{DeMiguel2023MachineAlpha}. These methods label funds but do not capture the generative process behind a strategy.

\subsection{Imitation Learning in Finance}

Imitation learning provides a framework for mimicking expert behavior from demonstrations, which aligns closely with our goal of cloning fund manager strategies without a predefined reward function. Research in this area has explored methods for inferring latent investment objectives. For example, \cite{Maeda2020LatentLearning} used multi-modal learning to segment traders based on their implicit goals. %Similarly,\cite{zhong2024soft} combined imitation learning with reinforcement learning to help trading agents generalize better to new market conditions.
These approaches have inspired our work, but they typically depend on some form of expert labeling or require a supplementary reward signal to guide the learning process. Our approach avoids this requirement by learning directly from the observed joint distribution of holdings and market data.

\subsection{Agent-Based Market Modeling}

Agent-based models (ABMs) study financial markets as complex adaptive systems emerging from individual agent interactions \cite{farmer2009economy,paulin2018agent}. ABM realism depends on diverse, authentic agent behaviors. Current approaches use hand-crafted utility functions or simplified rules that need significant effort to properly calibrate \cite{Platt2020AMethods}. For example, \cite{Scholl2021HowMalfunction} demonstrates how three stylized trading strategies (value investors, trend followers, and noise traders) interact in complex, density-dependent ways, but the model relies on manually specified strategy parameters and behaviors. Recent work has advanced market simulation:  \cite{Vyetrenko2019GetSimulations} developed realism metrics for limit order book simulations, while \cite{Yagi2020AnalysisSimulation} analyzed maker-taker fee impacts using agent-based simulation.

Our work directly addresses ABM calibration challenges by providing a data-driven method for learning diverse, empirically-grounded investment strategies from mutual fund data, enabling more realistic agent populations in market simulations.

%%%%%%%%%%%%%%%%%%%%%%%%%%%%%%%%%%%%%%%%%%%%%%%%%%%%%%%%%%%%%%%%%%%%%%%%%%%%%%%%%%%%%%%%%%%%%%%%%%%%
% 3 EXPERIMENTS
\section{Methods}\label{section:methods}

%% BREVITY
%This section details our generative framework for learning fund managers' strategies from public data. We define the problem as conditional probability estimation, describe our GAN-based architecture, and explain the training objective.

\subsection{Generative Problem Formulation}\label{section:problem_formulation}

We formulate fund strategy learning as estimating the conditional distribution:
\begin{equation}
    p_M\left(\mathbf{w}_{i,t} \mid \mathbf{w}_{i,t{-}1}, ~ \mathbf{X}_{i, t{-}1}, ~ \mathbf{r}_{i, t{-}T\ldots t{-}1},  ~ \boldsymbol{\phi}_{a,t}\right)
\end{equation}

where $\mathbf{w} \in \mathbb{R}^{N}$ represents portfolio weights, $\mathbf{X} \in \mathbb{R}^{N \times K}$ captures asset characteristics, $\mathbf{r} \in \mathbb{R}^{N \times T}$ contains historical returns for $t{-}T,\ldots,t{-}2,t{-}1$, and $\boldsymbol{\phi}_a \in \mathbb{R}^{d}$ encodes the latent strategy of manager $a$.

This formulation differs fundamentally from portfolio optimization-based approaches. Rather than assuming managers maximize some composite utility function of returns and risk and associated weights, %in essence $\max_{\mathbf{w}} \mathbb{E}[U(\mathbf{w}^T\mathbf{r})] $ subject to constraints, 
which not only are heterogeneous among fund managers, but are often not publicly declared and must be assumed, we learn the implicit mapping from market states to portfolio decisions probabilistically.
This means we model the distribution of plausible portfolios, that resemble a manager's previously observed allocations, conditional on market states.

\begin{figure}
    \centering
    \includegraphics[width=\linewidth]{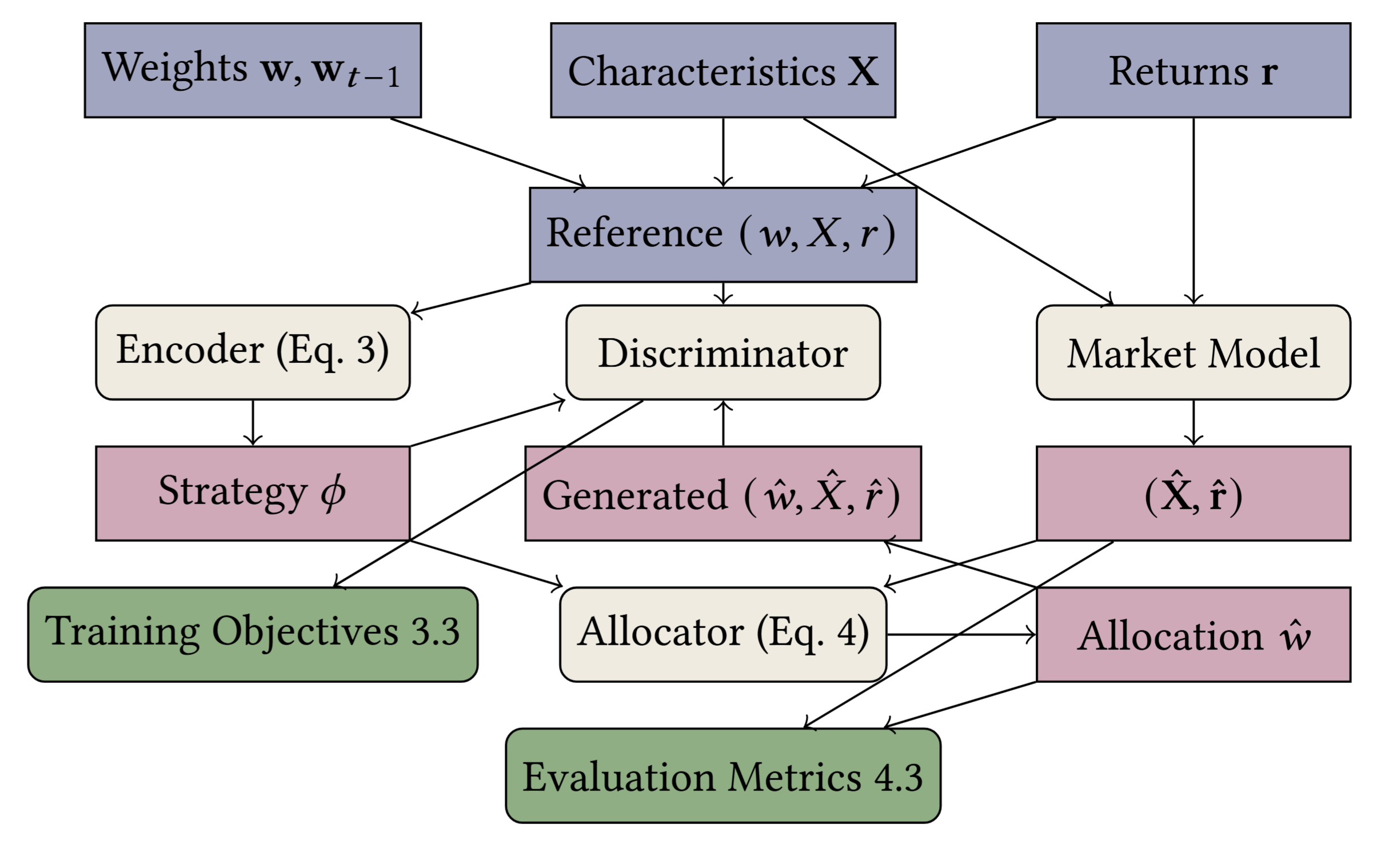}
    \caption{Architecture of the generative adversarial network for learning investment strategies. The framework consists of four main components: (1) A market model that generates synthetic stock universes $(\mathbf{\hat{X}}, \mathbf{\hat{r}})$ based on the Carhart four-factor model; (2) A strategy encoder that maps observed portfolio allocations to latent strategy representations $\boldsymbol{\phi}$' (3) A portfolio allocator (decoder) that generates realistic portfolio weights $\mathbf{\hat{w}}$ conditioned on market states and strategy encodings; and (4) A discriminator that distinguishes between real and generated portfolio-market data tuples. Blue boxes represent input data, red boxes show latent representations and synthetic recreations, beige nodes indicate neural network components, and green nodes denote training objectives and evaluation metrics.}
    \Description[A diagram showing a Neural Network Architecture.]{The figure contains a schematic representation of a neural network's major components.}
    \label{figure:architecture-overview}
% \end{figure*}
\end{figure}

\subsection{Architecture}\label{section:architecture}

Our architecture is a conditional Generative Adversarial Network (GAN). The diagram in Figure~\ref{figure:architecture-overview} summarizes the architecture. It has two components: a Generator that creates realistic portfolio allocations and a Discriminator that distinguishes generated portfolios from real ones. The investment universe contains thousands of stocks with numerous characteristics, creating a high-dimensional learning problem. We address this by using characteristic representations from a market model and an encoder-decoder generator structure.

The dimensionality of market states is large, with thousands of stocks with many characteristics each. The number of samples is small. Early experiments showed models trained only on the real data stock universe performed poorly out-of-sample. For this reason we use a two-stage generator that can be evaluated on synthetic stock universes.

\subsubsection{Generative Component for the Investment Universe}\label{section:universe_encoder}

The investment universe $\mathcal{U} = (X, r)$ is a tuple of characteristics and returns, for a collection of $N=500$ stocks in random order. $X\in \mathbb{R}^{N\times K}$ are $K$ characteristics, and $r\in \mathbb{R}^{N \times T}$ provides a history of $T$ periods of log returns. We choose $N= 500$ because most funds invest in 500 or fewer stocks, typically using the S\&P 500 as their benchmark and universe.

A stock universe generator $G_{\mathcal{U}}(X, r)$ creates synthetic market states. Rather than learning complex asset return dynamics from scratch, we embed the Carhart four-factor model \citep{Carhart1997OnPerformance} structure directly into our Variational Autoencoder architecture. The model posits that asset returns are explained by exposures $\beta$, obtained from a regression on four risk factors:

\begin{equation}\label{equation:factor_model}
r_{i,t} = \alpha_i + \sum_{k=1}^{4} \beta_{i,k} y_{k,t} + \epsilon_{i,t}
\end{equation}

where \(\alpha_i\) are idiosyncratic returns and \( \epsilon_{i,t} \) are the residuals of the regression.

% \begin{equation}\label{equation:factor_model}
% r_{i,t} = \alpha_i + \underbrace{\sum_{k=1}^{4} \beta_{i,k} y_{k,t}}_{\text{systematic return}} + \underbrace{\epsilon_{i,t}}_{\text{idiosyncratic return}}
% \end{equation}

The model provides good explanatory power while remaining parsimonious. It has features from three important categories: market, fundamental, and technical characteristics:
% \begin{itemize}
% \item[$y_1$] Market: Value-weighted market return minus risk-free rate
% \item[$y_2$] Size (SMB): Small minus big market capitalization
% \item[$y_3$] Value (HML): High minus low book-to-market ratio  
% \item[$y_4$] Momentum (UMD): Prior 12-month return winners minus losers
% \end{itemize}

\textbf{$y_1$} Market: Value-weighted market return minus risk-free rate

\textbf{$y_2$} Size (SMB): Small minus big market capitalization

\textbf{$y_3$} Value (HML): High minus low book-to-market ratio  

\textbf{$y_4$} Momentum (UMD): Prior 12-month return winners minus losers

% \noindent 
Pre-computed loadings are available from CRSP \cite{crsp_carhart_factors}, making the model and coefficients well-studied.

The encoder maps real market states $(X,r)$ to latent distributions. After solving the factor model (Eq.~\ref{equation:factor_model}), it uses attention mechanisms over dimensionally-reduced return cross-sections to learn low-dimensional representations of systematic factor shocks $y$ and residuals $\epsilon$. It separately encodes asset characteristics $X$ (containing factor loadings $\beta$ and intercepts $\alpha$). The output provides parameters (mean and log-variance) for the latent market state.

The decoder samples from the latent distribution to generate a synthetic market state $(\hat X, \hat r)$. Both components structurally enforce the factor model. The decoder first decodes the latent variable into synthetic characteristics $\hat X$ (containing $\hat \alpha, \hat \beta$) and factor shocks $\hat{y}$. The systematic portion of returns is then deterministically computed directly as $ \hat \beta \hat{y}^{\intercal}$. We further sample the idiosyncratic returns $\epsilon$, for which the decoder neural network learns idiosyncratic volatility. Finally, the returns with the desired structure are constructed by applying the factor model Equation~\ref{equation:factor_model}.

\subsubsection{Strategy Encoder}\label{section:strategy_encoder}

The strategy encoder $E_\phi$ maps observed portfolio allocations to a latent strategy representation $\boldsymbol{\phi} \in \mathbb{R}^8$. This encoder processes the complete portfolio context through three parallel lanes, one for each input tensor (characteristics, returns, weights), in order to preserve interpretability of how different data sources contribute to strategy representation.

\begin{equation}\label{equation:portfolio_manager_encoder}
    E_{\phi}(\mathbf{X}, \mathbf{r}, \mathbf{w}_{t{-}1}, \mathbf{w}_t) \rightarrow (\boldsymbol{\mu}_{\phi}, \log \boldsymbol{\sigma}^2_{\phi})
\end{equation}

The first lane processes characteristics $\mathbf{X}$ and weights $\mathbf{w}$, to capture how the portfolio tilts relative to market factors. It computes both portfolio-weighted and universe-average statistics, producing a 4-dimensional latent representation of factor exposures, $\phi_{1-4}$.

The second lane analyzes historical returns $\mathbf{r}$ weighted by portfolio allocations. It uses temporal reduction followed by attention mechanisms to capture return patterns and risk dynamics, outputting a 2-dimensional latent $\phi_{5-6}$ for performance characteristics.

The third lane examines weight changes $\mathbf{w}_t-\mathbf{w}_{t{-}1}$, to understand trading patterns and turnover behavior. This produces a 2-dimensional latent encoding the manager's trading aggressiveness and re-balancing style $\phi_{7-8}$.

Each lane employs the same architecture: after normalization and dimensionality reduction using dense layers, we use multi-head attention. The outputs are combined and processed through dense layers to produce the final 8-dimensional latent distribution parameters $(\boldsymbol{\mu}_{\phi}, \log \boldsymbol{\sigma}^2_{\phi})$.

\begin{figure*}%[t]%{0.48\columnwidth}
    \centering
    \includegraphics[width=\linewidth]{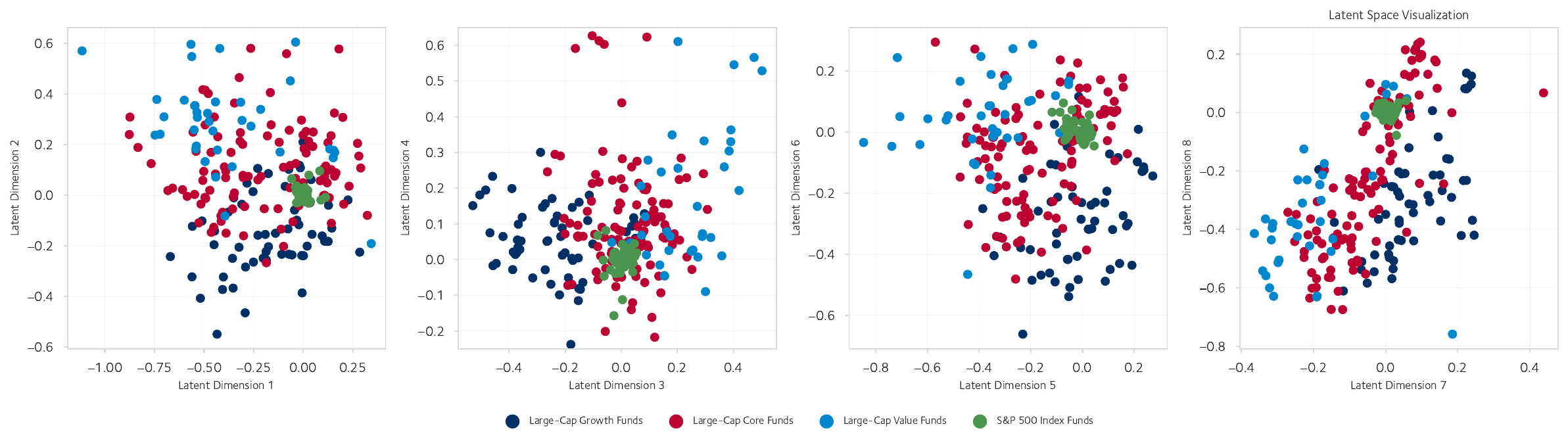}
    % \caption{This figure visualizes the latent variables for the strategy encoding $\phi_{a,t, 1-8}$ for the end of the first validation period,  $t=\text{December }2019$. Values are detrended, such that the S \& P 500 index lies at the origin. Recall from Section~\ref{}, latent variables $\phi_{1-4}$ are constructed from characteristics,   }
    \caption{Two-dimensional visualization of learned strategy representations in latent space for December 2020. Each point represents a fund's strategy encoding $\boldsymbol{\phi}_a$, with coordinates detrended relative to the S\&P 500 index (positioned at origin). The separation between groups of portfolios with the same label demonstrates that the learned representations successfully capture known investment styles without supervision.}\Description[Four scatter plots of latent spaces.]{Four scatter plots of latent spaces.}
    \label{figure:latent_space_visualization}
\end{figure*}

\subsubsection{Portfolio Allocator (Decoder)}
The portfolio allocator $D_w$ generates portfolio weights conditioned on the market state and latent strategy:
\begin{equation}\label{equation:portfolio_manager_decoder}
    D_w(\mathbf{X}, \mathbf{r}, \boldsymbol{\phi}, \mathbf{w}_{t{-}1}) \rightarrow \mathbf{\hat{w}}_t
\end{equation}

The allocator must produce valid portfolio allocations with appropriate sparsity while capturing the manager's strategy. It processes the latent strategy $\boldsymbol{\phi}$ alongside market information, to produce weights reflecting both systematic factor exposures and idiosyncratic selection.

\subsubsection{Discriminator}
The discriminator $D$ distinguishes observations of real portfolios from generated allocations based on the complete context. To allow the discriminator to investigate all possible metrics of a portfolio, we design it to examine the full empirical distributions $
(\mathbf{w}^T \mathbf{X}, \mathbf{w}^T \mathbf{r}, \mathbf{w}_t, \mathbf{w}_{t{-}1}, \boldsymbol{\phi})
$. % \begin{equation}
% D(\mathbf{X}, \mathbf{r}, \mathbf{w}, \mathbf{w}{t-1}, \boldsymbol{\phi}) = \text{NN}(\mathbf{w}^T \mathbf{X}, \mathbf{w}^T \mathbf{r}, \mathbf{w}, \mathbf{w}{t-1}, \boldsymbol{\phi})
% \end{equation}
% This design enables the discriminator to learn about not only the portfolio's weighted average features, but also other properties of the distribution.
This design enables the discriminator to learn portfolio weighted-average features and other distributional properties.
We employ a Wasserstein GAN \cite{Arjovsky2017WassersteinGAN} with gradient penalty (WGAN-GP) \cite{Gulrajani2017ImprovedGANs} for stable training. % 
% \begin{equation}\label{equation:wasserstein_loss}
% \mathcal{L}_D (Q_1, Q_2) = \mathbb{E}_{Q_1 \sim p_{\text{data}}}[D(Q_1)] - \mathbb{E}_{Q_2 \sim p{\text{generator}}}[D(Q_2)] + \lambda_{gp}\mathcal{L}_{gp}
% \end{equation}
The discriminator determines the Wasserstein distance between sample sets. When presented with real and synthetic data, we interpret this distance as a generator loss: larger distances mean the discriminator easily distinguishes real from synthetic samples. % The discriminator is designed to determine the Wasserstein distance between two sets of samples. When presenting the discriminator with real data and synthetic data, we can interpret this distance directly as a loss for the generator: the larger the distance, the easier it is for the discriminator to tell real and synthetic samples apart.% The discriminator architecture has a similar architecture as the strategy encoder, applying multi-head attention to a reduced form of the full portfolio.% This design gives us the opportunity to evaluate the generator on two related tasks:
Its architecture resembles the strategy encoder, applying multi-head attention to a reduced portfolio form. This design is sufficient to evaluate the generator on two related tasks:

\begin{multline}\label{equation:discriminator_tests}
\mathcal{L}_{\text{replication}} = \mathcal{L}_{D}\left(
\left(\mathbf{X}, \mathbf{r}, \mathbf{w}, \boldsymbol{\phi}\right), ~ \left(\mathbf{ X}, \mathbf{ r}, \mathbf{\hat w}, \boldsymbol{\phi}\right)\right), \\
\mathcal{L}_{\text{synthetic}} = \mathcal{L}_{D}\left(
\left(\mathbf{X}, \mathbf{r}, \mathbf{w}, \boldsymbol{\phi}\right), ~ \left(\mathbf{\hat X}, \mathbf{\hat r}, \mathbf{\hat w}, \boldsymbol{\phi}\right)\right).
\end{multline}
In this equation, the first loss $\mathcal{L}_{\text{replication}}$ measures the conditional portfolio allocation component's ability to replicate the original weights given the real market data. The second loss $\mathcal{L}_{\text{synthetic}}$ compares the distributions of real data against fully synthetic data comprising both the simulated stock universe and hypothetical allocation therein. This dual evaluation allows us to separately assess the quality of portfolio generation and the realism of the complete generative model.

\subsection{Training Objectives}\label{section:training_objectives}

We use the Wasserstein GAN with Gradient Penalty (WGAN-GP) objective for its training stability \cite{Gulrajani2017ImprovedGANs}. The Discriminator loss $\mathcal{L}_D$  follows the standard WGAN-GP formulation. The Generator's objective combines the adversarial loss with regularization terms to ensure financial realism while maintaining stability. The regularization terms also enable an ablation study where we evaluate the generator without discriminator feedback.
%The regularization terms will also serve an ablation test, where we disable the discriminator to see whether these terms alone are enough to generate realistic behavior. 

% The full generator loss $\mathcal{L}_G$ is:
% \begin{equation}
% \mathcal{L}_G = -\underset{\tilde{\mathbf{q}} \sim p_G}{\mathbb{E}}[D(\tilde{\mathbf{q}})] + \lambda_{1}\mathcal{L}{\text{exposure}} + \lambda_{2}\mathcal{L}{\text{concentration}} + \lambda_{3}\mathcal{L}{\text{turnover}}
% \end{equation}
% where the first term is the adversarial loss designed to fool the discriminator. 

% The regularization terms are $L_2$ penalties that encourage the generated portfolio $\hat{w}$ to match the real portfolio $w$ on key metrics:
% \begin{align}%\label{equation:regularization_metrics}
% \mathcal{L}{\text{exposure}} &= ( \mathbf{\hat{w}}^T \mathbf{X} - \mathbf{w}^T \mathbf{X} )^2 \label{equation:mean_exposure_regularization} \\
% \mathcal{L}{\text{concentration}} &= \left(\sum\mathbf{\hat{w}}^2 - \sum\mathbf{w}^2\right)^2 \label{equation:concentration_regularization} \\
% \mathcal{L}{\text{turnover}} &= (|\mathbf{\hat{w}}_{t} - \mathbf{w}_{t-1}|_1 - |\mathbf{w}_{t} - \mathbf{w}_{t-1}|_1)^2 \label{equation:turnover_regularization}
% \end{align}
% These terms ensure the generated portfolios match the target's factor tilts, concentration (Herfindahl Index), and turnover rate, respectively.

The full generator loss $\mathcal{L}_G$ is:
\begin{equation}
\mathcal{L}_G = \lambda_{1}\mathcal{L}_{\text{replication}} + \lambda_{2}\mathcal{L}_{\text{synthetic}} + \lambda_3 \mathcal{L}_{\text{exposure}}
\end{equation}
where all terms are adversarial losses. The $L_2$ regularization term ${L}_{\text{exposure}} = ( \mathbf{\hat{w}}^T \mathbf{X} - \mathbf{w}^T \mathbf{X} )^2$ encourages the generated portfolio $\hat{\mathbf{w}}$ to match the real portfolio $\mathbf{w}$ on the key metric of factor tilt.

% \begin{align}%\label{equation:regularization_metrics}
% \mathcal{L}_{\text{exposure}} &= ( \mathbf{\hat{w}}^T \mathbf{X} - \mathbf{w}^T \mathbf{X} )^2 \label{equation:mean_exposure_regularization} \\
% \mathcal{L}{\text{concentration}} &= \left(\sum\mathbf{\hat{w}}^2 - \sum\mathbf{w}^2\right)^2 \label{equation:concentration_regularization} \\
% \mathcal{L}{\text{turnover}} &= (|\mathbf{\hat{w}}_{t} - \mathbf{w}_{t-1}|_1 - |\mathbf{w}_{t} - \mathbf{w}_{t-1}|_1)^2 \label{equation:turnover_regularization}
% \end{align}
% These terms ensure the generated portfolios match the target's factor tilts, concentration (Herfindahl Index), and turnover rate, respectively.

\subsection{Implementation Details}

% We implement the strategy encoder, portfolio weight allocator and discriminator as deep neural networks. 
% We train using Adam optimizer with learning rates $10^{-4}$ (generator) and $10^{-4}$ (discriminator), updating the discriminator 3 times per generator update.

We use 8 latent dimensions, following the rule of thumb of one dimension per metric we can think of (4 Carhart factors, 2 for mean and variance parameters of returns, 2 more for turnover and concentration, based on weights). Our results suggest this number of latent dimensions provides more than sufficient capacity to capture information about the strategies, as several dimensions exhibit redundancy as there is significant cross-correlation in the latent dimensions. We think the dimensions of the latent representation could be reduced to favor parsimony, or the extra dimensionality can be exploited to shape the latent space in a way that is better interpretable. That is out of scope for this paper. We train using the Adam optimizer with learning rates $10^{-4}$ (generator) and $10^{-4}$ (discriminator), updating the discriminator 3 times per generator update.

%%%%%%%%%%%%%%%%%%%%%%%%%%%%%%%%%%%%%%%%%%%%%%%%%%%%%%%%%%%%%%%%%%%%%%%%%%%%%%%%%%%%%%%%%%%%%%%%%%%%
% 4 EXPERIMENTS
\section{Experiments}

% We design experiments to assess our framework's ability to reconstruct realistic portfolios, learn meaningful strategy representations, and discover implicit objectives beyond stated investment styles.
We design experiments to test our framework's ability to reconstruct realistic portfolios, learn meaningful strategy representations, and discover implicit objectives beyond stated investment styles.

\subsection{Data}
% We use the CRSP Survivor-Bias-Free US Mutual Fund Database, a standard dataset for academic research in finance. Our sample focuses on actively managed U.S. equity funds from 2010 to 2024. To ensure data quality, we apply several filters: funds must have at least 12 months of holdings data, report at least 75\% of their holdings by weight, and have at least 75\% of their allocation within our reference stock universe (the largest 500 U.S. stocks by market capitalization). After filtering, our final dataset comprises 1436 unique mutual funds with at least one year of observations for each. In total, we have over 120,000 observations of a portfolio at a point in time. For each fund, we use the Lipper classification (e.g., "Large-Cap Growth", "Small-Cap Value") as a ground-truth label.
We use the CRSP Survivor-Bias-Free US Mutual Fund Database, a standard academic finance dataset. Our sample covers actively managed U.S. equity funds from 2010 to 2024. We apply several quality filters: funds must have at least 12 months of holdings data, report at least 75\% of holdings by weight, and allocate at least 75\% within our stock universe (the largest 500 U.S. stocks by market cap). After filtering, we have 1436 unique mutual funds with at least one year of observations each. This gives us over 120,000 portfolio observations. We use Lipper classifications (e.g., "Large-Cap Growth", "Small-Cap Value") as ground-truth labels.
The data splits temporally: training on 2010-2018, validation on 2019, and testing on 2020-2024, ensuring no look-ahead bias.
% The data is split temporally: training on 2010-2018, validation to determine early stopping on 2019, and testing on 2020-2024, ensuring no look-ahead bias.

\subsection{Baselines}\label{section:baseline_definitions}
% We evaluate our full model against a series of increasingly less trivial baseline models and an ablation model to demonstrate the value of each component of our framework:

We evaluate our model against increasingly sophisticated baselines and an ablation study:

% \begin{enumerate}

    % \item \textbf{Zero-Trade (ZT):} A simple but strong baseline that maintains previous period weights. The portfolio for the current period is identical to the previous period's portfolio, reflecting a pure buy-and-hold strategy: $\mathbf{\hat{w}}_t = \mathbf{w}_{t-1}$. 
    % \item 
    (1) \textbf{Zero-Trade (ZT):} Maintains previous period weights as a buy-and-hold strategy: $\mathbf{\hat{w}}_t = \mathbf{w}_{t{-}1}$. This simple baseline is surprisingly strong.
    % \item \textbf{Turnover-Matched Random (TMR):} This baseline tests whether portfolio allocations can be explained by random trading alone, while respecting a manager's typical trading rate. It generates a new portfolio by adding a random perturbation to the previous weights, $\mathbf{w}_{t-1}$. The magnitude of this perturbation is scaled such that the L1-norm of the trade, $\| \mathbf{\hat{w}}_t - \mathbf{w}_{t-1} \|_1$, matches the historical turnover of the target fund. This isolates the effect of strategic selection from the effect of trading frequency.
    % \item 
    (2) \textbf{Turnover-Matched Random (TMR):} Tests whether allocations can be explained by random trading at the manager's typical rate. It adds random perturbations to previous weights, scaling the magnitude so turnover $| \mathbf{\hat{w}}_t - \mathbf{w}_{t{-}1} |_1$ matches the target fund's historical rate. This isolates strategic selection from trading frequency.

    % \item \textbf{Factor-Tilt Matched (FTM):} This advanced baseline argues that a manager's primary skill lies in maintaining a target exposure to risk factors. It first generates a portfolio using the TMR method. It then iteratively adjusts this portfolio through rejection sampling until its factor exposures (tilts), $\mathbf{\hat{w}}_t^T \mathbf{X}$, are within a small tolerance of 10\% of the target fund's actual factor exposures, $\mathbf{w}_t^T \mathbf{X}$. This model controls for both turnover and explicit factor tilting.
    % \item 
    (3) \textbf{Factor-Tilt Matched (FTM):} Argues that managers primarily maintain target factor exposures. It generates portfolios using TMR, then adjusts through rejection sampling until factor exposures $\mathbf{\hat{w}}_t^T \mathbf{X}$ are within 10\% of the target's actual exposures $\mathbf{w}_t^T \mathbf{X}$. This controls for both turnover and explicit factor tilting.

    % \item \textbf{Generator-Only (Ablation):} To isolate the contribution of the adversarial training process, we evaluate our generator model trained without the discriminator. The generator is trained using only the reconstruction-based regularization losses ($\mathcal{L}_{\text{exposure}}$, $\mathcal{L}_{\text{concentration}}$, $\mathcal{L}_{\text{turnover}}$) from Equations 6-8. This tests whether simply matching explicit portfolio characteristics is sufficient to generate realistic portfolios.
    % \item 
    (4) \textbf{Generator-Only (Ablation):} Our generator trained without the discriminator, using only reconstruction losses ($\mathcal{L}{\text{exposure}}$, $\mathcal{L}{\text{concentration}}$, $\mathcal{L}_{\text{turnover}}$). This tests whether matching explicit characteristics suffices for realistic portfolios.
% \end{enumerate}

\subsection{Evaluation Metrics}\label{section:evaluation_metrics}

% A fundamental challenge in evaluating fund manager models stems from the absence of formal utility specifications. Real-world portfolio construction involves proprietary processes and undisclosed objectives that extend far beyond simple risk-return optimization. We cannot write a comprehensive objective function that captures all aspects of manager behavior because the true objectives remain unknown.
% This presents a methodological challenge: while we could add numerous loss components through weighted combinations, choosing these weights becomes arbitrary and doesn't address the core issue of hidden objectives. Moreover, we cannot determine whether the discriminator identifies meaningful portfolio characteristics or merely exploits superficial modeling deficiencies.
% To address this challenge, we evaluate our model using metrics deliberately excluded from training. This approach allows us to assess whether the model captures genuine investment behaviors rather than overfitting to our chosen training objectives.

A fundamental challenge in evaluating fund manager models stems from the absence of formal utility specifications. Real-world portfolio construction involves proprietary processes and undisclosed objectives that extend far beyond simple risk-return optimization. We cannot write a comprehensive objective function capturing all manager behaviors because the heterogeneous true objectives remain unknown.
This creates a methodological challenge: while we could add numerous loss components through weighted combinations, choosing weights becomes arbitrary and doesn't address hidden objectives. We cannot determine whether the discriminator identifies meaningful characteristics or exploits modeling deficiencies.
We address this by evaluating using metrics deliberately excluded from training. This tests whether the model captures genuine investment behaviors that we know to exist, rather than overfitting to misspecified or incorrectly weighted training objectives.

\subsubsection{Portfolio Reconstruction Quality}\label{section:portfolio_reconstruction}
% We measure the error between a generated portfolio $\mathbf{\hat{w}}$ and the ground-truth portfolio $\mathbf{w}$ using metrics that were \textit{not} explicitly part of the regularization loss function, ensuring a fair evaluation.
We measure error between generated portfolio $\mathbf{\hat{w}}$ and ground-truth $\mathbf{w}$ using metrics not in the regularization loss:

% \begin{itemize}
%     \item \textbf{Count Error ($\mathcal{L}_{\text{count}}$):} The absolute difference in the number of assets held (with weight $>$ 0.01\%). Tests realistic portfolio sparsity.
%     \item \textbf{Concentration Error ($\mathcal{L}_{\text{concentration}}$):} The absolute difference between the Herfindahl indices of the generated and real portfolios, $\text{abs}( \|\mathbf{\hat{w}}\|_2^2 - \|\mathbf{w}\|_2^2)$. This measures the accuracy of portfolio concentration.
%     \item \textbf{Turnover Error ($\mathcal{L}_{\text{turnover}}$):} The squared difference in portfolio turnover, $\text{abs}((\|\mathbf{\hat{w}}_{t} - \mathbf{w}_{t{-}1}\|_1 - \|\mathbf{w}_{t} - \mathbf{w}_{t{-}1}\|_1)$. This tests if the model captures the manager's trading aggressiveness.
% \end{itemize}

 \textbf{Count Error ($\mathcal{L}_{\text{count}}$):} The absolute difference in the number of assets held (with weight $>$ 0.01\%). Tests realistic portfolio sparsity.

 \textbf{Concentration Error ($\mathcal{L}_{\text{concentration}}$):} The absolute difference between the Herfindahl indices of the generated and real portfolios, $\text{abs}( \|\mathbf{\hat{w}}\|_2^2 - \|\mathbf{w}\|_2^2)$. This measures the accuracy of portfolio concentration.

 \textbf{Turnover Error ($\mathcal{L}_{\text{turnover}}$):} The squared difference in portfolio turnover, $\text{abs}((\|\mathbf{\hat{w}}_{t} - \mathbf{w}_{t{-}1}\|_1 - \|\mathbf{w}_{t} - \mathbf{w}_{t{-}1}\|_1)$. This tests if the model captures the manager's trading aggressiveness.

\subsubsection{Strategy Representation Quality}\label{section:strategy_representation}

% A good model should learn a latent space $\boldsymbol{\phi}$ where strategies are meaningfully organized. We test this by probing the latent space with a linear classifier (an SVM). We train the SVM to predict the expert-provided Lipper classification labels from the learned strategy embeddings $\boldsymbol{\phi}_a$. The key metric is the macro-averaged recall, which measures the average per-class accuracy and is robust to class imbalance. A high score indicates that the latent space linearly separates real-world investment styles.
A good model should organize strategies meaningfully in latent space $\boldsymbol{\phi}$. We probe this with a linear classifier (SVM) to recover Lipper classifications from learned embeddings $\boldsymbol{\phi}_a$. The key metric is macro-averaged recall, measuring average per-class accuracy while being robust to class imbalance. High scores indicate the latent space linearly separates real investment styles.
\paragraph{Classification}\label{section:classification}
% To verify our latent space captures meaningful strategies, we probe whether it linearly separates known investment styles. The Lipper classification scheme serves as our benchmark expert model, which categorizes funds by averaging their factor exposures over a 36-month rolling window.
We verify our latent space captures meaningful strategies by testing linear separation of known investment styles. The Lipper scheme categorizes funds by averaging factor exposures over 36-month rolling windows to mitigate market noise, we do the same with the latent representation $\boldsymbol{\phi}_a
$ for each portfolio $a$. We then train a linear SVM on these averaged embeddings to predict Lipper classifications. Using simple linear classifiers tests whether the latent space contains expert knowledge in an easily interpretable form.
% We evaluate classification performance using macro-averaged recall. For each category, recall is computed as $\text{TP} / (\text{TP} + \text{FN})$, measuring the fraction of true fund members correctly identified. These per-category scores are then averaged across all categories, giving equal weight regardless of class size. This metric is particularly appropriate for our imbalanced dataset where some fund styles are more common than others. High macro-averaged recall indicates the model discovers financially meaningful representations without supervision.
We use macro-averaged recall for evaluation. For each category, recall is $\text{TP} / (\text{TP} + \text{FN})$, measuring the fraction of true members correctly identified. These scores are averaged across categories, giving equal weight regardless of class size. This suits our imbalanced dataset where some styles are more common. High macro-averaged recall indicates the model discovers financially meaningful representations without supervision.

\subsubsection{Behavioral Fidelity}\label{section:behavioral_fidelity}
We design two tests for subtle, implicit behaviors:

% \textbf{Strategy Stability:} Real fund managers maintain a consistent style over time. We measure the stability of a strategy by calculating the drift of its factor tilts over time, relative to the market average. We define the drift for fund $a$ as $u_a = \frac{1}{T}\sum_{t=1}^{T} \| (\boldsymbol{\beta}_{a,t} - \bar{\boldsymbol{\beta}}_t) - (\boldsymbol{\beta}_{a,t-1} - \bar{\boldsymbol{\beta}}_{t-1}) \|_1$, where $\boldsymbol{\beta}_{a,t} = \mathbf{w}_{a,t}^T \mathbf{X}$ are the fund's factor tilts. Lower drift implies higher stability.
\textbf{Strategy Stability:} Real managers are constrained by the prospectus to maintain consistent styles over time. We measure stability by calculating factor tilt drift relative to market average. For fund $a$: $u_a = \frac{1}{T}\sum_{t=1}^{T} | (\boldsymbol{\beta}_{a,t} - \bar{\boldsymbol{\beta}}_t) - (\boldsymbol{\beta}_{a,t{-}1} - \bar{\boldsymbol{\beta}}_{t{-}1}) |_1$, where $\boldsymbol{\beta}_{a,t} = \mathbf{w}_{a,t}^T \mathbf{X}$ are factor tilts. Lower drift implies higher stability.

% \textbf{Markowitz Optimal-Proximity:} We test the hypothesis that many managers implicitly behave like Markowitz optimizers, even if not explicitly stated. For each time period, we construct the \textit{ex-post} efficient frontier based on the realized returns and covariance of the asset universe. We then calculate the Euclidean distance of each fund's portfolio from this frontier in risk-return space. A smaller distance suggests the portfolio was closer to being optimal. We report the percentage of funds in each Lipper category that are closer to the efficient frontier than a market-cap weighted benchmark portfolio.
\textbf{Markowitz Optimal-Proximity:} We test whether managers implicitly behave like Markowitz optimizers. There are many models of optimality, but this is well-known and easy to test. For each period, we construct the ex-post efficient frontier from realized returns and covariance. We calculate each fund's distance from this frontier in risk-return space and rank it compared to a sample of style-matched random portfolios. Smaller distances compared to random suggest a higher likelihood that the portfolio is optimized. We report the average proximity score per Lipper class.

\subsubsection{Counterfactual Analysis}\label{section:counterfactual_analysis}

% To verify that learned strategies encode transferable investment principles rather than mere memorization of test cases, we conduct comprehensive counterfactual experiments. These tests examine whether a strategy learned in one market context can be meaningfully applied to different conditions while preserving its essential characteristics. We extract latent strategies $\boldsymbol{\phi}_a$ from portfolios operating in one time period and apply them to generate portfolios in different market regimes, testing whether the fundamental investment approach remains consistent.
We verify that learned strategies encode transferable investment principles rather than memorization through comprehensive counterfactual experiments. These tests examine whether strategies learned in one market context apply meaningfully to different conditions while preserving essential characteristics. We extract latent strategies $\boldsymbol{\phi}_a$ from portfolios in one time period and apply them to different market regimes, including completely novel synthetic environments, testing whether fundamental investment approaches remain consistent.

% Our analysis focuses on three key aspects of strategy transfer. First, we verify that factor exposures—the systematic tilts toward value, growth, momentum, and size—remain stable when strategies are applied to new market conditions. This tests whether the model captures persistent investment philosophies rather than temporary market positions. Second, we examine portfolio concentration and turnover patterns to ensure the generated portfolios maintain realistic structural properties across different contexts. Finally, we perform bidirectional strategy swaps between time periods to confirm that relative differences between strategies (e.g., growth versus value orientations) are preserved regardless of market regime. Together, these experiments demonstrate that our framework learns robust strategy representations that generalize beyond their training contexts.

Our analysis focuses on three aspects of strategy transfer. First, we verify that factor exposures—systematic tilts toward Value and Growth remain stable when applied to new market conditions. This tests whether the model captures persistent investment philosophies rather than temporary positions. Second, we examine portfolio concentration and turnover patterns to ensure generated portfolios maintain realistic structural properties across contexts. Finally, we perform bidirectional strategy swaps between time periods to confirm that relative differences between strategies are preserved regardless of market regime, which also informs us about the stability of the classification over time.

\begin{table*}[ht]
\caption{Comparative performance across training objectives and hold-out evaluation metrics (test set average).  Baseline models are specified in Section~\ref{section:baseline_definitions}. Training objectives include replication loss (portfolio reconstruction on real data), synthetic loss (generation on synthetic universes), and overall generator loss. Hold-out metrics assess portfolio realism on characteristics deliberately excluded from training (\ref{section:evaluation_metrics}).}
\label{table:reconstruction_per_model}
\begin{tabular}{lrrr|rrr}
\toprule
\multicolumn{1}{c}{} & \multicolumn{3}{c|}{\textbf{Training Objectives}} & \multicolumn{3}{c}{\textbf{Hold-Out Metrics}} \\
\cmidrule(lr){2-4} \cmidrule(lr){5-7}
Model & $\mathcal{L}_{\text{replication}}$ & $\mathcal{L}_{\text{synthetic}}$ & $\mathcal{L}_{\text{generator}}$ & $\mathcal{L}_{\text{count}}$ & $\mathcal{L}_{\text{concentration}}$ & $\mathcal{L}_{\text{turnover}}$ \\
\midrule
1. Zero-Trade                   &  0.063 & 0.830                     & 2.817           & 23          & 0.0072            & 0.1716 \\
2. Random Trade                 &  0.068 & 0.831                     & 2.810           & 34          & 0.0095            & \textbf{0.0415} \\
3. Factor-Tilt                  &  0.144 & 0.506                     & 3.003           & 31          & 0.0064            & 1.2776 \\
4. Generator-Only               &  0.201 & 0.263                     & 2.820           & 84          & 0.0089            & 0.9232 \\
5. Full GAN                     &  \textbf{0.061} & \textbf{0.236}   & \textbf{1.0882} & \textbf{15} & \textbf{0.0047}   & 0.5451 \\ % 
\bottomrule
\end{tabular}
\end{table*}

% \pagebreak % COSTMETIC HINT
\paragraph{Experimental Design}\label{section:experimental_design}
% We implement four complementary tests to assess the quality of strategy transfer:
We implement the following tests of strategy transfer quality:
% \begin{enumerate}
    % \item
    
(1) \textbf{Strategy Swap Test}: We extract the latent strategy $\boldsymbol{\phi}_1$ from one portfolio operating in universe $\mathcal{U}_1$ and apply it to a different universe $\mathcal{U}_2$. This tests whether strategies encode transferable allocation principles rather than universe-specific memorization.
    
    % \item 
(2) \textbf{Strategy Preservation Test}: We compare factor exposures between original and counterfactual portfolios to verify that key investment characteristics are maintained during transfer.
% \item \textbf{Portfolio Concentration Analysis}: We measure the number of holdings and Herfindahl concentration index to ensure realistic portfolio construction in counterfactual scenarios.
% \item \textbf{Strategy Transfer Quality}: We perform bidirectional strategy swaps to verify that relative differences between strategies are preserved across universes.
% \end{enumerate}

% \paragraph{Implementation Details}\label{section:implementation_detauls}

% For each test, we sample portfolios from all time periods in the test set to ensure distinct market conditions. To handle mixed precision training artifacts, we implement dtype conversion to ensure consistent float32 operations throughout the analysis.

%%%%%%%%%%%%%%%%%%%%%%%%%%%%%%%%%%%%%%%%%%%%%%%%%%%%%%%%%%%%%%%%%%%%%%%%%%%%%%%%%%%%%%%%%%%%%%%%%%%%
% 5 RESULTS
\section{Results}

We evaluate our framework across reconstruction quality, strategy representation, and behavioral fidelity metrics. The full GAN architecture outperforms baselines on most metrics, demonstrating the value of adversarial training for learning realistic investment strategies.

% \subsection{Wasserstein Distance}

\subsection{Reconstruction Quality}

Table \ref{table:reconstruction_per_model} presents reconstruction metrics across all models. The discriminator maintains positive Wasserstein distances even at convergence which is a common outcome, confirming slight discriminator dominance rather than theoretical equilibrium . Despite this, the full model achieves the lowest errors on key hold-out metrics: count error (15 stocks), concentration matching (0.0047). 
%  with the standard 3:1 critic-to-generator update ratio—

The Zero-Trade baseline performs surprisingly well on concentration, replication metrics, reflecting the slow turnover in fund holdings. Indeed, the average turnover in our sample ranges between 100\%-250\% per year. The Generator-Only ablation test shows degraded performance across many metrics except the performance on the synthetic data, with count error increasing to 84 stocks, demonstrating that adversarial feedback is essential for realistic portfolio generation beyond the basic statistical moments of the exposure. That is, the Generator-Only model fails to be realistic because it finds it easier to achieve the desired exposures through allocating to a large number of different stocks. As expected, the random trades with matching turnover score best on the turnover metrics.
% Table \ref{table:reconstruction_per_model} presents reconstruction metrics across all models. The discriminator maintains positive Wasserstein distances even at convergence which is a common outcome with the standard 3:1 critic-to-generator update ratio confirming slight discriminator dominance rather than theoretical equilibrium. Despite this, the full model achieves the lowest errors on key metrics: count error (15 stocks), concentration matching (0.0047), and turnover fidelity (0.5451).

% Table \ref{table:reconstruction_per_model} presents reconstruction metrics across all models. The discriminator maintains positive Wasserstein distances even at convergence—a common outcome with the standard 3:1 critic-to-generator update ratio—confirming slight discriminator dominance rather than theoretical equilibrium. Despite this, the full model achieves the lowest errors on key metrics: count error (15 stocks), concentration matching (?????), and turnover fidelity (????). The Zero-Trade baseline performs surprisingly well on concentration and replication metrics, reflecting conservation in fund holdings. However, the synthetic data shows less persistence in risk factor loadings, causing a static strategy to rapidly diverge for the completely novel stock universe. The Generator-Only ablation shows degraded performance across all metrics, with count error increasing to 84 stocks, demonstrating that adversarial feedback is essential for realistic portfolio generation.

\subsection{Classification Performance}

Linear probing of the 8-dimensional latent space achieves a macro-averaged recall of 77\% when predicting Lipper classes. Brief experimentation reveals this can be increased to 95\% with a non-linear SVM kernel, suggesting that the bulk of the Lipper classes is embedded in the latent space in a straightforward linear way, and that if increased congruence is desired this could be achieved by transformation of the latent space.  Figure \ref{figure:latent_space_visualization} visualizes the learned representations for December 2020, with coordinates detrended relative to the S\&P 500 index. Growth and value funds form distinct clusters in the latent space, while core funds occupy intermediate positions. Dimensions 1-4, constructed from factor exposures, show the clearest separation between investment styles, raising our confidence the strategy encoder is using risk factor exposures to model different investment styles.

\begin{table}
\caption{Classification performance for investment style prediction using learned latent representations. Linear SVM trained on strategy encodings $\boldsymbol{\phi}$ to recovered Lipper fund categories. A macro-averaged score of 77\% shows the Lipper scheme is contained in the latent space in a simple linear form.  }
\begin{tabular}{lllll}
\toprule
                               & Precision & Recall & F1 Score & Support \\ 
\midrule
Large Cap Core            & 0.81      & 0.88   & 0.84     & 214     \\
Large Cap Growth          & 0.88      & 0.83   & 0.85     & 138     \\
Large Cap Value           & 0.68      & 0.44   & 0.53     & 39      \\
S\&P 500 Index            & 0.86      & 0.94   & 0.90     & 32      \\
% accuracy                       & 0.83      & 423    &          &         \\
Macro Average                      & 0.81      & 0.77   & 0.78     & 423     \\
% weighted avg                   & 0.82      & 0.83   & 0.82     & 423     \\ 
\bottomrule
\end{tabular}
\end{table}

\subsubsection{Latent Space Interpretation}

We analyze correlations between the 8-dimensional latent space and portfolio characteristics including Carhart factor loadings, Sharpe ratio, allocation size, and turnover rate.
The latent dimensions show moderate but interpretable correlations with known factors. Dimensions $\phi_1$ and $\phi_7$ strongly correlate with SMB factor loading and Sharpe ratio, capturing small-cap exposure and performance. Dimension $\phi_2$ correlates most strongly with HML, encoding value-growth orientation. Dimensions $\phi_4$ and $\phi_6$ correlate with allocation size, reflecting portfolio concentration. 
The moderate correlation strengths indicate the latent space captures more than simple (linear) factor combinations in each dimension. The apparent redundancy across multiple dimensions likely reflects the model's adaptation to noisy financial data, where redundant but similar representations built from different inputs can provide additional consistency in noisy tasks.

\subsection{Behavioral Consistency Metrics}

\subsubsection{Strategy Stability}
Real fund strategies exhibit an average drift of $u=0.13$ in factor space units, compared to $u=1.34$ for the unconstrained random trading baseline. This metric remains consistent between the Generator-Only model and the full architecture, suggesting the exposure regularization alone suffices for capturing strategy persistence, as the generator on its own achieves the desired mean exposures. When trading randomly but constrained by investment style (FTM), the drift is roughly twice that of the real funds and one-fifth of purely random trading, suggesting that the latent space is strongly laid out to reflect average exposures.

% gen only (0.131), discr (0.133)

% ???
% PLOT OF DRIFT OF DIFFERENT FUNDS
% ???

\subsubsection{Proximity to Efficient Frontier}
Table \ref{tab:optimization} reports the average proximity score for real portfolios and portfolios generated by the models. Among real funds, 95.5\% of Index funds and 90.4\% of Growth funds show evidence of mean-variance optimization compared to simple random portfolios, while only 67.0\% of Value funds exhibit this pattern. The same test repeated for the portfolios generated by the two models produces different percentages. The Generator-Only model does not achieve the same level of optimization, while the full mode does slightly better. This is an area for improvement, and we find that it can be improved by explicit utility modeling by including the Sharpe ratio as an objective. This remains out of scope for this paper.

\begin{table}[h]
    \caption{This table shows the proximity scores for different classes of portfolios.}
    \label{tab:optimization}
    \begin{tabular}{lrrr}
    \toprule
    Fund Category & Empirical & Generator Only & Full Model \\
    \midrule
    S\&P 500 Index    &  95.5\% & 61.6\% & 91.1\% \\
    Growth (all caps) &  90.4\% & 49.8\% & 48.7\% \\
    Core (all caps)   &  82.7\% & 37.7\% & 58.2\% \\
    Value (all caps)  &  67.0\% & 37.4\% & 52.5\% \\
    \bottomrule
    \end{tabular}
\end{table}

\subsection{Counterfactual Results}\label{section:counterfactual_results}

We evaluate strategy transferability by extracting latent representations from one market context and applying them to generate portfolios in different universes. Table \ref{table:counterfactual_results} summarizes the key metrics.
The full GAN model demonstrates robust strategy transfer. When swapping strategies between different market universes, the generated portfolios maintain factor exposures close to the original strategy. For instance, the SMB exposure changes minimally from -0.773 to -0.767, while the market beta shifts from -1.037 to -0.835—preserving the fund's defensive stance. 
In contrast, the Generator-Only model struggles with strategy transfer. Factor exposures deviate significantly: SMB exposure collapses from -0.773 to -0.164, and momentum (UMD) drops from 0.171 to 0.042. 
We combine these insights with results form Table~\ref{table:reconstruction_per_model}. Most tellingly, the model generates portfolios with 189 holdings—nearly double the original—with a Herfindahl index of 0.0089, indicating excessive diversification. This confirms our earlier finding that, without adversarial feedback, the generator achieves target exposures through unrealistic over-diversification.

\begin{table}
\caption{This table summarizes strategy transfer across different market universes. $\Delta$ values show the absolute change in factor exposures between original and transferred strategies. The full GAN model demonstrates robust strategy transfer with minimal drift in key factors, while the Generator-Only model shows significant degradation. These results validate that adversarial training learns transferable investment principles.}
\label{table:counterfactual_results}
\begin{tabular}{lcc}
    \toprule
    Metric & Full GAN & Generator-Only \\
    \midrule
    % \multicolumn{3}{l}{\textit{Factor Exposure Preservation}} \\
    $\Delta$ Market Beta & 0.202 & 0.438 \\
    $\Delta$ SMB         & 0.006 & 0.609 \\
    $\Delta$ HML         & 0.055 & 0.052 \\
    $\Delta$ UMD         & 0.098 & 0.129 \\
    \bottomrule
\end{tabular}
\end{table}
These results validate that the adversarial training is necessary not just for realistic portfolio generation, but for learning transferable strategy representations. The discriminator forces the generator to capture the essence of investment strategies—their systematic tilts and concentration preferences—rather than merely matching statistical moments through any available means.
    % \midrule
    % \multicolumn{3}{l}{\textit{Portfolio Structure}} \\
    % Holdings (Original) & 105 & 105 \\
    % Holdings (Counterfactual) & 139 & 189 \\
    % Herfindahl (Original) & 0.0229 & 0.0229 \\
    % Herfindahl (Counterfactual) & 0.0157 & 0.0089 \\
    % \midrule
    % % \multicolumn{3}{l}{\textit{Strategy Transfer}} \\
    % Transfer Quality Ratio & 1.49 & 1.88 \\

\section{Discussion}\label{section:discussion}

Our generative framework successfully learns fund manager strategies without explicit utility specification. The results reveal several important findings. First, adversarial training proves essential—the discriminator forces the generator beyond simple moment matching to capture the full distributional properties of real portfolios. Second, the learned 8-dimensional latent space meaningfully organizes investment strategies, with linear separability of expert classifications achieving high macro-averaged recall. Third, our analysis uncovers implicit optimization behaviors: over four-fifths of active funds exhibit Markowitz-like efficiency seeking.
The framework faces unique challenges inherent to financial data. Very few holdings-level data is publicly available. Market regimes shift over time, making stationarity assumptions problematic for long-term predictions of behavior. Our temporal data splitting prevents look-ahead bias but may not capture sufficient variety of regimes. Additionally, monthly holdings snapshots and the relatively short history of high-quality data (post-2010) limit our ability to model higher-frequency trading strategies or validate performance across multiple market cycles.

\subsection{Beyond Utility Functions}\label{section:beyond_utility}

Our results demonstrate the practical value of learning complex fund strategies without explicit utility specification. The learned representations capture interactions between multiple competing objectives as per the prospectus (style adherence) and unstated constraints (turnover limits, tracking error), as these funds form a continuum in representations.
The partitioned latent space reveals that while factor exposures explain significant variance (dimensions $\phi_1$-$\phi_4$), additional dimensions capturing return patterns and turnover are necessary for realistic strategy generation. This supports the hypothesis that managers behave heterogeneously, optimizing or trading off multiple potentially conflicting objectives, with various degrees of success.
This approach offers particular advantages for agent-based market modeling, where specifying realistic utility functions for diverse agents has been a persistent challenge. Rather than hand-tuning utility parameters (which often results in unrealistic agent behaviors), our framework enables sampling from the empirical distribution of real fund strategies. This data-driven approach can improve the realism of market simulations by populating them with agents whose behaviors are grounded in observed fund management practices.

\subsection{Implications for Market Simulation}\label{section:market_simulation}

Our framework directly addresses a fundamental challenge in agent-based market modeling: the specification of realistic, heterogeneous agent behaviors. Current ABMs rely on hand-crafted utility functions or simplified trading rules that may poorly approximate real market participants. By learning strategies directly from fund data, we enable a new approach for ABM construction. We can now populate simulations with agents whose behaviors are empirically grounded rather than theoretically postulated. The learned strategy representations can be sampled to create diverse agent populations, interpolated to explore strategy variations, or perturbed to study market dynamics under counterfactual scenarios. This data-driven approach promises more realistic emergent market properties and better calibration to empirical stylized facts.

\subsection{Limitations and Future Work}\label{section:limitations}

%%% BREVITY
%Several limitations inform future research directions. 
Our monthly holdings data cannot capture higher-frequency components of a strategy, potentially missing important market dynamics. The focus on U.S. equity funds, while providing clean data, excludes multi-asset strategies, derivatives usage, and international markets. The 2010-2024 training period, does miss such interesting periods as the 2008 crisis, and may not generalize to different market regimes or future conditions.
Future work should explore conditional generation based on market regimes, investigating robustness to changing conditions. Extending to multi-asset portfolios would require modeling cross-asset dependencies. Integration with market microstructure models could bridge the gap between monthly holdings and daily price formation. Real-time applications might use the framework for strategy monitoring, anomaly detection, or regulatory oversight. Finally, theoretical work connecting our empirical findings to economic models of bounded rationality and multiple objectives could deepen our understanding of investment behavior.

\section{Conclusions}

We present a generative adversarial framework for learning investment strategies from fund holdings data without assuming explicit utility specifications. Our key finding is that fund manager behavior exhibits fundamental heterogeneity that is difficult to capture at scale using traditional utility-based models.
The learned representations reveal a continuous spectrum of strategies rather than discrete clusters. Even within expert-defined categories we observe variation of behaviors. The Lipper classification's hard boundaries impose artificial discretization. This heterogeneity extends beyond style factors to multiple behavioral dimensions: the degree of mean-variance optimization varies across funds. This heterogeneity explains why explicit utility specification approaches face fundamental challenges. The generative approach sidesteps these specification challenges by learning the complete decision process from observed behavior. Rather than requiring researchers to identify all relevant objectives and estimate their relative importance, our framework captures the full complexity of manager preferences through the learned strategy distributions.  This data-driven approach represents a key advantage over methods that require explicit specification of all relevant objectives. %Our framework achieves strong empirical performance across multiple evaluation metrics while remaining agnostic to the underlying utility structure.
While many aspects of the current model can be improved, we think that over the long run leveraging computation and data beats human domain knowledge and engineered features. Therefore this work opens new directions for practical applications where realistic behavior is needed but explicit specification is challenging. Agent-based market simulations, risk management systems, and regulatory monitoring can benefit from models that capture the full diversity of real investment strategies without requiring explicit parameterization of manager objectives. By learning what fund managers actually do rather than assuming what theory suggests they should do, we can model market participants more realistically.

% \paragraph
% \vspace{1em}
% {\textbf{ACKNOWLEDGMENTS}} Withheld for anonymous review.
\begin{acks}
This paper was prepared for informational purposes in part by the Artificial Intelligence Research group of JPMorgan Chase \& Co and its affiliates (``JP Morgan”), and is not a product of the Research Department of JP Morgan. JP Morgan makes no representation and warranty whatsoever and disclaims all liability, for the completeness, accuracy or reliability of the information contained herein. This document is not intended as investment research or investment advice, or a recommendation, offer or solicitation for the purchase or sale of any security, financial instrument, financial product or service, or to be used in any way for evaluating the merits of participating in any transaction, and shall not constitute a solicitation under any jurisdiction or to any person, if such solicitation under such jurisdiction or to such person would be unlawful. AC acknowledges funding from a UKRI AI World Leading Researcher Fellowship (grant EP/W002949/1) and from a JPMC Faculty Research Award.
\end{acks}

\bibliographystyle{ACM-Reference-Format}
\bibliography{references}

%%% -*-BibTeX-*-
%%% Do NOT edit. File created by BibTeX with style
%%% ACM-Reference-Format-Journals [18-Jan-2012].

\begin{thebibliography}{24}

%%% ====================================================================
%%% NOTE TO THE USER: you can override these defaults by providing
%%% customized versions of any of these macros before the \bibliography
%%% command.  Each of them MUST provide its own final punctuation,
%%% except for \shownote{} and \showURL{}.  The latter two
%%% do not use final punctuation, in order to avoid confusing it with
%%% the Web address.
%%%
%%% To suppress output of a particular field, define its macro to expand
%%% to an empty string, or better, \unskip, like this:
%%%
%%% \newcommand{\showURL}[1]{\unskip}   % LaTeX syntax
%%%
%%% \def \showURL #1{\unskip}           % plain TeX syntax
%%%
%%% ====================================================================

\ifx \showCODEN    \undefined \def \showCODEN     #1{\unskip}     \fi
\ifx \showISBNx    \undefined \def \showISBNx     #1{\unskip}     \fi
\ifx \showISBNxiii \undefined \def \showISBNxiii  #1{\unskip}     \fi
\ifx \showISSN     \undefined \def \showISSN      #1{\unskip}     \fi
\ifx \showLCCN     \undefined \def \showLCCN      #1{\unskip}     \fi
\ifx \shownote     \undefined \def \shownote      #1{#1}          \fi
\ifx \showarticletitle \undefined \def \showarticletitle #1{#1}   \fi
\ifx \showURL      \undefined \def \showURL       {\relax}        \fi
% The following commands are used for tagged output and should be
% invisible to TeX
\providecommand\bibfield[2]{#2}
\providecommand\bibinfo[2]{#2}
\providecommand\natexlab[1]{#1}
\providecommand\showeprint[2][]{arXiv:#2}

\bibitem[Arjovsky et~al\mbox{.}(2017)]%
        {Arjovsky2017WassersteinGAN}
\bibfield{author}{\bibinfo{person}{Martin Arjovsky}, \bibinfo{person}{Soumith
  Chintala}, {and} \bibinfo{person}{Léon Bottou}.}
  \bibinfo{year}{2017}\natexlab{}.
\newblock \showarticletitle{{Wasserstein GAN}}.
\newblock \bibinfo{journal}{\emph{Proceedings of the 34th International
  Conference on Machine Learning}}  \bibinfo{volume}{70} (\bibinfo{date}{1}
  \bibinfo{year}{2017}), \bibinfo{pages}{214--223}.
\newblock
\urldef\tempurl%
\url{https://arxiv.org/abs/1701.07875v3}
\showURL{%
\tempurl}


\bibitem[Brown et~al\mbox{.}(2009)]%
        {Brown2009StayingFunds}
\bibfield{author}{\bibinfo{person}{Keith~C. Brown}, \bibinfo{person}{W.~Van
  Harlow}, {and} \bibinfo{person}{Hanjiang Zhang}.}
  \bibinfo{year}{2009}\natexlab{}.
\newblock \bibinfo{booktitle}{\emph{{Staying the Course: The Role of Investment
  Style Consistency in the Performance of Mutual Funds}}}.
\newblock \bibinfo{type}{{T}echnical {R}eport}.
  \bibinfo{institution}{University of Texas at Austin}.
\newblock
\href{https://doi.org/10.2139/SSRN.1364737}{doi:\nolinkurl{10.2139/SSRN.1364737}}


\bibitem[Carhart(1997)]%
        {Carhart1997OnPerformance}
\bibfield{author}{\bibinfo{person}{Mark~M. Carhart}.}
  \bibinfo{year}{1997}\natexlab{}.
\newblock \showarticletitle{{On Persistence in Mutual Fund Performance}}.
\newblock \bibinfo{journal}{\emph{The Journal of Finance}}
  \bibinfo{volume}{52}, \bibinfo{number}{1} (\bibinfo{date}{3}
  \bibinfo{year}{1997}), \bibinfo{pages}{57--82}.
\newblock
\showISSN{00221082}
\href{https://doi.org/10.1111/j.1540-6261.1997.tb03808.x}{doi:\nolinkurl{10.1111/j.1540-6261.1997.tb03808.x}}


\bibitem[{Center for Research in Security Prices}(2024)]%
        {crsp_carhart_factors}
\bibfield{author}{\bibinfo{person}{{Center for Research in Security Prices}}.}
  \bibinfo{year}{2024}\natexlab{}.
\newblock \bibinfo{title}{{CRSP/Compustat Merged Database: Carhart Four-Factor
  Model Loadings}}.
\newblock \bibinfo{howpublished}{Wharton Research Data Services (WRDS)}.
\newblock
\urldef\tempurl%
\url{https://wrds-www.wharton.upenn.edu/}
\showURL{%
\tempurl}
\newblock
\shownote{Accessed: [2024]}.


\bibitem[Cont et~al\mbox{.}(2025)]%
        {Cont2022TailGAN}
\bibfield{author}{\bibinfo{person}{Rama Cont}, \bibinfo{person}{Mihai
  Cucuringu}, \bibinfo{person}{Renyuan Xu}, {and} \bibinfo{person}{Chao
  Zhang}.} \bibinfo{year}{2025}\natexlab{}.
\newblock \showarticletitle{Tail-GAN: Learning to Simulate Tail Risk
  Scenarios}.
\newblock \bibinfo{journal}{\emph{Management Science}} \bibinfo{volume}{0},
  \bibinfo{number}{0} (\bibinfo{year}{2025}), \bibinfo{pages}{1}.
\newblock
\href{https://doi.org/10.1287/mnsc.2023.00936}{doi:\nolinkurl{10.1287/mnsc.2023.00936}}
\newblock
\shownote{Articles in Advance}.


\bibitem[DeMiguel et~al\mbox{.}(2023)]%
        {DeMiguel2023MachineAlpha}
\bibfield{author}{\bibinfo{person}{Victor DeMiguel}, \bibinfo{person}{Javier
  Gil-Bazo}, \bibinfo{person}{Francisco~J. Nogales}, {and}
  \bibinfo{person}{André~A.P. Santos}.} \bibinfo{year}{2023}\natexlab{}.
\newblock \showarticletitle{{Machine learning and fund characteristics help to
  select mutual funds with positive alpha}}.
\newblock \bibinfo{journal}{\emph{Journal of Financial Economics}}
  \bibinfo{volume}{150}, \bibinfo{number}{3} (\bibinfo{date}{12}
  \bibinfo{year}{2023}), \bibinfo{pages}{103737}.
\newblock
\showISSN{0304-405X}
\href{https://doi.org/10.1016/J.JFINECO.2023.103737}{doi:\nolinkurl{10.1016/J.JFINECO.2023.103737}}


\bibitem[Farmer and Foley(2009)]%
        {farmer2009economy}
\bibfield{author}{\bibinfo{person}{J~Doyne Farmer} {and}
  \bibinfo{person}{Duncan Foley}.} \bibinfo{year}{2009}\natexlab{}.
\newblock \showarticletitle{The economy needs agent-based modelling}.
\newblock \bibinfo{journal}{\emph{Nature}} \bibinfo{volume}{460},
  \bibinfo{number}{7256} (\bibinfo{year}{2009}), \bibinfo{pages}{685--686}.
\newblock
\urldef\tempurl%
\url{https://doi.org/10.1038/460685a}
\showURL{%
\tempurl}


\bibitem[Gopal(2024)]%
        {Gopal2024NeuralFactors}
\bibfield{author}{\bibinfo{person}{Achintya Gopal}.}
  \bibinfo{year}{2024}\natexlab{}.
\newblock \showarticletitle{NeuralFactors: A Novel Factor Learning Approach to
  Generative Modeling of Equities}. In \bibinfo{booktitle}{\emph{Proceedings of
  the 5th ACM International Conference on AI in Finance}}
  \emph{(\bibinfo{series}{ICAIF '24})}. \bibinfo{publisher}{Association for
  Computing Machinery}, \bibinfo{address}{New York, NY, USA},
  \bibinfo{pages}{99--107}.
\newblock
\href{https://doi.org/10.1145/3677052.3698647}{doi:\nolinkurl{10.1145/3677052.3698647}}


\bibitem[Gulrajani et~al\mbox{.}(2017)]%
        {Gulrajani2017ImprovedGANs}
\bibfield{author}{\bibinfo{person}{Ishaan Gulrajani}, \bibinfo{person}{Faruk
  Ahmed}, \bibinfo{person}{Martin Arjovsky}, \bibinfo{person}{Vincent
  Dumoulin}, {and} \bibinfo{person}{Aaron Courville}.}
  \bibinfo{year}{2017}\natexlab{}.
\newblock \showarticletitle{{Improved Training of Wasserstein GANs}}.
\newblock \bibinfo{journal}{\emph{Advances in Neural Information Processing
  Systems}}  \bibinfo{volume}{2017-December} (\bibinfo{date}{3}
  \bibinfo{year}{2017}), \bibinfo{pages}{5768--5778}.
\newblock
\showISSN{10495258}
\urldef\tempurl%
\url{https://arxiv.org/abs/1704.00028v3}
\showURL{%
\tempurl}


\bibitem[Hasanhodzic and Lo(2007)]%
        {Hasanhodzic2007CanCase}
\bibfield{author}{\bibinfo{person}{Jasmina Hasanhodzic} {and}
  \bibinfo{person}{Andrew~W Lo}.} \bibinfo{year}{2007}\natexlab{}.
\newblock \showarticletitle{{Can Hedge-Fund Returns Be Replicated?: The Linear
  Case}}.
\newblock \bibinfo{journal}{\emph{Journal of Investment Management}}
  \bibinfo{volume}{5}, \bibinfo{number}{2} (\bibinfo{year}{2007}),
  \bibinfo{pages}{5--45}.
\newblock


\bibitem[Kaniel et~al\mbox{.}(2023)]%
        {Kaniel2023Machine-learningManagers}
\bibfield{author}{\bibinfo{person}{Ron Kaniel}, \bibinfo{person}{Zihan Lin},
  \bibinfo{person}{Markus Pelger}, {and} \bibinfo{person}{Stijn
  Van~Nieuwerburgh}.} \bibinfo{year}{2023}\natexlab{}.
\newblock \showarticletitle{{Machine-learning the skill of mutual fund
  managers}}.
\newblock \bibinfo{journal}{\emph{Journal of Financial Economics}}
  \bibinfo{volume}{150}, \bibinfo{number}{1} (\bibinfo{date}{10}
  \bibinfo{year}{2023}), \bibinfo{pages}{94--138}.
\newblock
\showISSN{0304-405X}
\href{https://doi.org/10.1016/J.JFINECO.2023.07.004}{doi:\nolinkurl{10.1016/J.JFINECO.2023.07.004}}


\bibitem[Kwon and Lee(2024)]%
        {Kwon2024CanSeries}
\bibfield{author}{\bibinfo{person}{Sohyeon Kwon} {and} \bibinfo{person}{Yongjae
  Lee}.} \bibinfo{year}{2024}\natexlab{}.
\newblock \showarticletitle{{Can GANs Learn the Stylized Facts of Financial
  Time Series?}}
\newblock \bibinfo{journal}{\emph{ICAIF 2024 - 5th ACM International Conference
  on AI in Finance}}  \bibinfo{volume}{1} (\bibinfo{date}{10}
  \bibinfo{year}{2024}), \bibinfo{pages}{126--133}.
\newblock
\showISBNx{9798400710810}
\href{https://doi.org/10.1145/3677052.3698661}{doi:\nolinkurl{10.1145/3677052.3698661}}


\bibitem[Maeda et~al\mbox{.}(2020)]%
        {Maeda2020LatentLearning}
\bibfield{author}{\bibinfo{person}{Iwao Maeda}, \bibinfo{person}{David deGraw},
  \bibinfo{person}{Michiharu Kitano}, \bibinfo{person}{Hiroyasu Matsushima},
  \bibinfo{person}{Kiyoshi Izumi}, \bibinfo{person}{Hiroki Sakaji}, {and}
  \bibinfo{person}{Atsuo Kato}.} \bibinfo{year}{2020}\natexlab{}.
\newblock \showarticletitle{{Latent Segmentation of Stock Trading Strategies
  Using Multi-Modal Imitation Learning}}.
\newblock \bibinfo{journal}{\emph{Journal of Risk and Financial Management
  2020, Vol. 13, Page 250}} \bibinfo{volume}{13}, \bibinfo{number}{11}
  (\bibinfo{date}{10} \bibinfo{year}{2020}), \bibinfo{pages}{250}.
\newblock
\showISSN{1911-8074}
\href{https://doi.org/10.3390/JRFM13110250}{doi:\nolinkurl{10.3390/JRFM13110250}}


\bibitem[Markowitz(1952)]%
        {Markowitz1952}
\bibfield{author}{\bibinfo{person}{Harry Markowitz}.}
  \bibinfo{year}{1952}\natexlab{}.
\newblock \showarticletitle{{Portfolio Selection}}.
\newblock \bibinfo{journal}{\emph{The Journal of Finance}} \bibinfo{volume}{7},
  \bibinfo{number}{1} (\bibinfo{year}{1952}), \bibinfo{pages}{77--91}.
\newblock
\urldef\tempurl%
\url{https://www.jstor.org/stable/2975974}
\showURL{%
\tempurl}


\bibitem[Paulin et~al\mbox{.}(2018)]%
        {paulin2018agent}
\bibfield{author}{\bibinfo{person}{James Paulin}, \bibinfo{person}{Anisoara
  Calinescu}, {and} \bibinfo{person}{Michael Wooldridge}.}
  \bibinfo{year}{2018}\natexlab{}.
\newblock \showarticletitle{Agent-based modeling for complex financial
  systems}.
\newblock \bibinfo{journal}{\emph{IEEE Intelligent Systems}}
  \bibinfo{volume}{33}, \bibinfo{number}{2} (\bibinfo{year}{2018}),
  \bibinfo{pages}{74--82}.
\newblock


\bibitem[Platt(2020)]%
        {Platt2020AMethods}
\bibfield{author}{\bibinfo{person}{Donovan Platt}.}
  \bibinfo{year}{2020}\natexlab{}.
\newblock \showarticletitle{{A comparison of economic agent-based model
  calibration methods}}.
\newblock \bibinfo{journal}{\emph{Journal of Economic Dynamics and Control}}
  \bibinfo{volume}{113} (\bibinfo{date}{4} \bibinfo{year}{2020}),
  \bibinfo{pages}{103859}.
\newblock
\showISSN{0165-1889}
\href{https://doi.org/10.1016/J.JEDC.2020.103859}{doi:\nolinkurl{10.1016/J.JEDC.2020.103859}}


\bibitem[Ramirez et~al\mbox{.}(2023)]%
        {Ramirez2023AAllocation}
\bibfield{author}{\bibinfo{person}{Domingo Ramirez},
  \bibinfo{person}{Jose~Manuel Pe{\~{n}}a}, \bibinfo{person}{Fernando
  Su{\'{a}}rez}, \bibinfo{person}{Omar Larr{\'{e}}}, {and}
  \bibinfo{person}{Arturo Cifuentes}.} \bibinfo{year}{2023}\natexlab{}.
\newblock \showarticletitle{{A Machine Learning Plus-Features Based Approach
  for Optimal Asset Allocation}}.
\newblock \bibinfo{journal}{\emph{ICAIF 2023 - 4th ACM International Conference
  on AI in Finance}} (\bibinfo{date}{11} \bibinfo{year}{2023}),
  \bibinfo{pages}{549--556}.
\newblock
\showISBNx{9798400702402}
\href{https://doi.org/10.1145/3604237.3626865}{doi:\nolinkurl{10.1145/3604237.3626865}}


\bibitem[Scholl et~al\mbox{.}(2021)]%
        {Scholl2021HowMalfunction}
\bibfield{author}{\bibinfo{person}{Maarten~P. Scholl},
  \bibinfo{person}{Anisoara Calinescu}, {and} \bibinfo{person}{J.~Doyne
  Farmer}.} \bibinfo{year}{2021}\natexlab{}.
\newblock \showarticletitle{How market ecology explains market malfunction}.
\newblock \bibinfo{journal}{\emph{Proceedings of the National Academy of
  Sciences}} \bibinfo{volume}{118}, \bibinfo{number}{26}
  (\bibinfo{year}{2021}), \bibinfo{pages}{e2015574118}.
\newblock
\href{https://doi.org/10.1073/pnas.2015574118}{doi:\nolinkurl{10.1073/pnas.2015574118}}


\bibitem[Sharpe(1966)]%
        {Sharpe1966MutualPerformance}
\bibfield{author}{\bibinfo{person}{William~F. Sharpe}.}
  \bibinfo{year}{1966}\natexlab{}.
\newblock \showarticletitle{{Mutual Fund Performance}}.
\newblock \bibinfo{journal}{\emph{The Journal of Business}}
  \bibinfo{volume}{39}, \bibinfo{number}{1} (\bibinfo{date}{1}
  \bibinfo{year}{1966}), \bibinfo{pages}{119--138}.
\newblock
\urldef\tempurl%
\url{https://www.jstor.org/stable/2351741}
\showURL{%
\tempurl}


\bibitem[Sharpe(1992)]%
        {Sharpe1992AssetAllocation}
\bibfield{author}{\bibinfo{person}{William~F. Sharpe}.}
  \bibinfo{year}{1992}\natexlab{}.
\newblock \showarticletitle{{Asset allocation}}.
\newblock \bibinfo{journal}{\emph{The Journal of Portfolio Management}}
  \bibinfo{volume}{18}, \bibinfo{number}{2} (\bibinfo{date}{1}
  \bibinfo{year}{1992}), \bibinfo{pages}{7--19}.
\newblock
\showISSN{0095-4918}
\href{https://doi.org/10.3905/JPM.1992.409394}{doi:\nolinkurl{10.3905/JPM.1992.409394}}


\bibitem[Vyetrenko et~al\mbox{.}(2019)]%
        {Vyetrenko2019GetSimulations}
\bibfield{author}{\bibinfo{person}{Svitlana Vyetrenko}, \bibinfo{person}{David
  Byrd}, \bibinfo{person}{Nick Petosa}, \bibinfo{person}{Mahmoud Mahfouz},
  \bibinfo{person}{Danial Dervovic}, \bibinfo{person}{Manuela Veloso}, {and}
  \bibinfo{person}{Tucker Balch}.} \bibinfo{year}{2019}\natexlab{}.
\newblock \showarticletitle{{Get Real: Realism Metrics for Robust Limit Order
  Book Market Simulations}}.
\newblock \bibinfo{journal}{\emph{ICAIF 2020 - 1st ACM International Conference
  on AI in Finance}} (\bibinfo{date}{12} \bibinfo{year}{2019}).
\newblock
\showISBNx{9781450375849}
\href{https://doi.org/10.1145/3383455.3422561}{doi:\nolinkurl{10.1145/3383455.3422561}}


\bibitem[Wiese et~al\mbox{.}(2019)]%
        {Wiese2019QuantSeries}
\bibfield{author}{\bibinfo{person}{Magnus Wiese}, \bibinfo{person}{Robert
  Knobloch}, \bibinfo{person}{Ralf Korn}, {and} \bibinfo{person}{Peter
  Kretschmer}.} \bibinfo{year}{2019}\natexlab{}.
\newblock \showarticletitle{{Quant GANs: Deep Generation of Financial Time
  Series}}.
\newblock \bibinfo{journal}{\emph{Quantitative Finance}} \bibinfo{volume}{20},
  \bibinfo{number}{9} (\bibinfo{date}{12} \bibinfo{year}{2019}),
  \bibinfo{pages}{1419--1440}.
\newblock
\href{https://doi.org/10.1080/14697688.2020.1730426}{doi:\nolinkurl{10.1080/14697688.2020.1730426}}


\bibitem[Xu et~al\mbox{.}(2019)]%
        {Xu2019ModelingGAN}
\bibfield{author}{\bibinfo{person}{Lei Xu}, \bibinfo{person}{Maria
  Skoularidou}, \bibinfo{person}{Alfredo Cuesta-Infante}, {and}
  \bibinfo{person}{Kalyan Veeramachaneni}.} \bibinfo{year}{2019}\natexlab{}.
\newblock \showarticletitle{{Modeling Tabular data using Conditional GAN}}.
\newblock \bibinfo{journal}{\emph{Advances in Neural Information Processing
  Systems}}  \bibinfo{volume}{32} (\bibinfo{date}{7} \bibinfo{year}{2019}).
\newblock
\showISSN{10495258}
\urldef\tempurl%
\url{https://arxiv.org/pdf/1907.00503}
\showURL{%
\tempurl}


\bibitem[Yagi et~al\mbox{.}(2020)]%
        {Yagi2020AnalysisSimulation}
\bibfield{author}{\bibinfo{person}{Isao Yagi}, \bibinfo{person}{Mahiro
  Hoshino}, {and} \bibinfo{person}{Takanobu Mizuta}.}
  \bibinfo{year}{2020}\natexlab{}.
\newblock \showarticletitle{{Analysis of the impact of maker-taker fees on the
  stock market using agent-based simulation}}.
\newblock \bibinfo{journal}{\emph{ICAIF 2020 - 1st ACM International Conference
  on AI in Finance}} (\bibinfo{date}{10} \bibinfo{year}{2020}).
\newblock
\showISBNx{9781450375849}
\href{https://doi.org/10.1145/3383455.3422523}{doi:\nolinkurl{10.1145/3383455.3422523}}


\end{thebibliography}
% \bibliography{all_references}

\end{document}